\documentclass[preprint,12pt]{elsarticle}
\usepackage{graphics}
\usepackage{subfigure}
\usepackage{graphicx}
\usepackage{amsmath}
\usepackage{float}

\usepackage{amssymb}

\journal{Physica A}

\begin{document}

\begin{frontmatter}



\title{The average velocity of self-propelled particles in a two-dimensional potential with colored noise}

\author{Bing Wang}
\author{Xiuqing Wu}
\author{Changzhao Chen}
\author{Mengjun Hu}
\author{Xuanyan Cao}
\address{Department of Physics and Mathematics, Anhui University of Science and Technology, Huainan, 232001, P.R.China}

\begin{abstract}
The average velocity of self-propelled particles in a two-dimensional potential with colored noise is investigated. The current reversal phenomenon appear with changing $x$ direction colored noise intensity. There exist optimal values of the parameters at which the average velocity takes its maximal value. The $y$ direction noise and the self-propelled angle noise have great effects on the $x$ direction average velocity, but they can not induce $x$ direction particles transport phenomenon themselves.
\end{abstract}


\end{frontmatter}

\section{\label{label1}Introduction}
The transport of particles induced by zero average fluctuations and random perturbations plays a crucial role in many physical and biological systems, which has many theoretical and practical implications\cite{b.a1, b.a2, b.a3, b.a4}. Some Brownian ratchets have been proposed to model the unidirectional motion driven by  nonequilibrium fluctuations and random perturbations, such as diffusion ratchets, rocking ratchets, flashing ratchets, and so on. These previous ratchet models mainly focus on passive Brownian particles. Recently, works of active Brownian particles have received some attention\cite{b.b1, b.b2}. Self-propelled particles have been investigated as the simplest model for "active materials", and the kinetic of self-propelled particles moving in potential could exhibit peculiar behavior\cite{b.c1, b.c2, b.c3, b.c4, b.c5, b.c6}.

There are numerous realizations of self-propelled particles in nature ranging from bacteria and spermatozoa to artificial colloidal microswimmers. D'Orsogna \textsl{et al.} modeled a self-propelling biological or artificial individuals interacting through pairwise attractive and repulsive forces, and predicted stability and morphology of organization starting from the shape of the two-body interaction\cite{b.c1}. H. Chat\'{e} \textsl{et al.} investigated Vicsek-style self-propelled particle models and observed a statistically homogeneous ordered phase\cite{b.c3}. Angelani \textsl{et al.} studied the run and tumble particles in periodic potential and found the asymmetric potential produces a net drift speed\cite{b.d1, b.d2}. Wan \textsl{et al.} showed a rectification phenomenon which is similar to experimental results achieved in a simple model for over damped swimming bacteria\cite{b.e1}. Ghosh  \textsl{et al.} studied the transport of Janus particles in periodically compartmentalized channel and found that the rectification can be orders of magnitude stronger than that for ordinary thermal potential ratchets\cite{b.e2}. Potosky \textsl{et al.} found that even in a symmetric potential a spatially modulated self-propelled velocity can induce the directed transport\cite{b.f}. Hagen \textsl{et al.}  studied the over damped Brownian motion of a self-propelled particle by solving the Langevin equation analytically and calculated the first four moments of the probability distribution function\cite{b.g1}. In all these studies, the potential is quasi-one-dimensional and independent in each dimension.

In 2014, Ai \textsl{et al.} investigated the rectification and diffusion of self-propelled particles in a two-dimensional corrugated channel with white noise, and found the self-propelled velocity can strongly increase the effective diffusion, while the large rotational diffusion rate can strongly suppress the effective diffusion\cite{b.i}. In this paper, we extended these studies to the case of a two-dimensional potential with colored noise. The paper is organized as follows: In Section \ref{label2}, the basic model of self-propelled ratchets with a two-dimensional potential and colored noise is provided. In Section \ref{label3}, the effects of parameters is investigated by means of simulations.  In Section \ref{label4}, we get the conclusions.

\section{\label{label2}Basic model and methods}
\begin{figure}
\center{
\includegraphics[height=6cm,width=8cm]{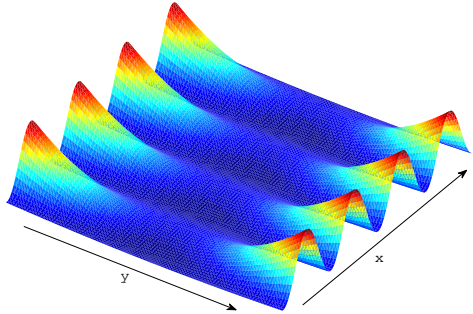}
\caption{The two dimensional periodic potential U(x,y), which is periodic in x direction and parabolic in y direction.} \label{Fig0}}
\end{figure}

In the present work, we attempt to consider self-propelled particles that move in a two dimensional potential. The dynamics of Brownian particle is governed by the following Langevin equations\cite{b.h1,b.h2}.
\begin{equation}
\frac{dx}{dt}=v_0\cos\theta+\mu F_x+\xi_1(t), \label{xt}
\end{equation}
\begin{equation}
\frac{dy}{dt}=v_0\sin\theta+\mu F_y+\xi_2(t), \label{yt}
\end{equation}
\begin{equation}
\frac{d\theta}{dt}=\xi_3(t), \label{zt}
\end{equation}
$v_0$ is the self-propelled velocity and $\mu$ is the mobility. $\theta$ is the self-propelled angle. $F_x=-\frac{\partial{U(x,y)}}{\partial{x}}$ and $F_y=-\frac{\partial{U(x,y)}}{\partial{y}}$.
\begin{equation}
U(x,y)=-U_0\sin(\frac{2\pi x}{L})+\frac{1}{2}C_0[1-\lambda \sin( \frac{2\pi x}{L}+\phi)]y^2,
\end{equation}
is the potential, which periodic in $x$ direction and parabolic in $y$ direction\cite{b.h3}. $U_0$ is the height of the $x$ direction potential. $C_0$ is the intensity of the $y$ direction potential. $\phi$ is the phase shift between the $x$ direction potential and the modulation function. $\lambda$ is the modulation constant with $0<\lambda<1$. $\xi_1$ is the $x$ direction Gaussian colored noise. $\xi_2$ is the $y$ direction Gaussian colored noise. $\xi_3$ is the self-propelled angle colored noise, and describes the nonequilibrium angular fluctuation. $\xi_{1,2,3}$  satisfies the following relations:
\begin{equation}
\langle\xi_i(t)\rangle=0,
\end{equation}
\begin{equation}
\langle\xi_i(t)\xi_j(s)\rangle=\delta_{ij}\frac{Q_i}{\tau_i}\exp[-\frac{|t-s|}{\tau_i}], i,j=1,2,3,
\end{equation}
$\langle\cdots\rangle$ denotes an ensemble average over the distribution of the random forces. $Q_1$($Q_2$) is the $x$($y$) direction colored noise intensity. $Q_3$ is the colored noise intensity of self-propelled angle. $\tau_1$, $\tau_2$ and $\tau_3$ are the self-correlation time of the noises, respectively. Upon introducing characteristic length scale $L$, time scale $\tau_0=\frac{L^2}{\mu U_0}$, and energy $U_0$ , Eqs. (\ref{xt}, \ref{yt}, \ref{zt}) can be rewritten in dimensionless form:
\begin{equation}
\frac{d\hat{x}}{d\hat{t}}=\hat{v}_0\cos\theta+\hat{F}_{\hat{x}}+\hat{\xi}_{1}(\hat{t}), \label{hxt}
\end{equation}
\begin{equation}
\frac{d\hat{y}}{d\hat{t}}=\hat{v}_0\sin\theta+\hat{F}_{\hat{y}}+\hat{\xi}_{2}(\hat{t}), \label{hyt}
\end{equation}
\begin{equation}
\frac{d\theta}{d\hat{t}}=\hat{\xi}_{3}(\hat{t}), \label{hzt}
\end{equation}
with $\hat{x}=\frac{x}{L}$, $\hat{y}=\frac{y}{L}$, $\hat{t}=\frac{t}{\tau_0}$, $\hat{U}=\frac{U}{U_0}$, $\hat{v}_0=\frac{v_0L}{\mu U_0}$, $\hat{Q}_1 =\frac{Q_1}{\mu U_0}$, $\hat{Q}_2 =\frac{Q_2}{\mu U_0}$, $\hat{Q}_3 =\frac{Q_3L^2}{\mu U_0}$, $\hat{\tau}_1=\frac{\tau_1}{\tau_0}$, $\hat{\tau}_2=\frac{\tau_2}{\tau_0}$, $\hat{\tau}_3=\frac{\tau_3}{\tau_0}$.
 The  dimensionless form of potential(show in Fig.(\ref{Fig0})) is rewritten as
 \begin{equation}
\hat{U}(\hat{x},\hat{y})=-\sin(2\pi \hat{x})+\frac{1}{2}\hat{C}_0[1-\lambda \sin( 2\pi \hat{x}+\phi)]\hat{y}^2
\end{equation}
with $\hat{C}_0 =\frac{C_0L^2}{U_0}$. In this paper, we will use only the dimensionless Eqs. (\ref{hxt},\ref{hyt},\ref{hzt}) and shall omit the hat($\wedge$) for all quantities\cite{b.i}.

A central practical question in the theory of Brownian motors is the over all long-time behavior of the particle, and the key quantities of particle transport through periodic potential is the particle velocity $V$.  $V$ can be corroborated by Brownian dynamic simulations performed by integration of the Langevin equations using the stochastic Euler algorithm. We only calculate the $x$ direction average velocity as the potential along $y$ direction is parabolic. The $x$ direction average velocity can be obtained from the following equation:
 \begin{equation}
V=\lim_{t\to\infty}\frac{\langle{x(t)-x_0}\rangle}{t-t_0}
\end{equation}
$x(t)$ is the position of particles at time $t$,  and $x(t_0)=x_0$.

\section{\label{label3}Results and discussion}
In order to give a simple and clear analysis of the system, Eqs.(\ref{hxt}), (\ref{hyt}) and (\ref{hzt}) are integrated using the Euler algorithm with $C_0=5.0$, $\phi=\frac{\pi}{2}$ and time step $\Delta t=10^{-4}$. The total integration time was more than $1.5\times10^6$.  The stochastic averages reported above were obtained as ensemble averages over $3\times 10^4$ trajectories with random initial conditions.

\begin{figure}
\center{
\includegraphics[height=6cm,width=8cm]{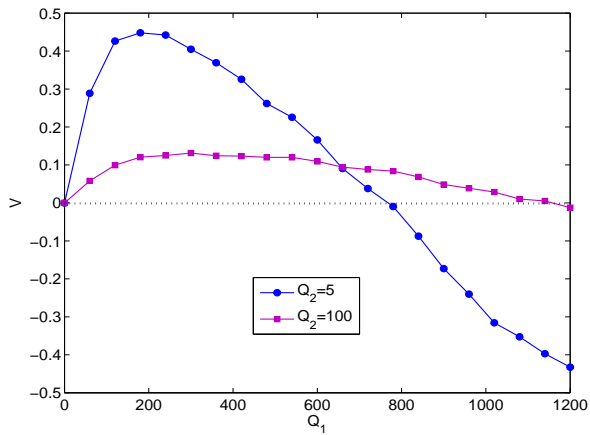}
\caption{Average velocity $V$ as a function of the $x$ direction noise intensity $Q_1$ with different $Q_2$. The other parameters are $Q_3=0.01$, $\lambda=0.9$, $v_0=2$ and $\tau_1=\tau_2=\tau_3=1$.} \label{Fig2}}
\end{figure}
The average velocity $V$ as a function of the $x$ direction noise intensity $Q_1$ is reported in Fig.\ref{Fig2}. We find that there exist a maximum with increasing $Q_1$, which shows the feature of resonance. The smaller $y$ direction noise intensity $Q_2$, the resonance phenomenon is more obvious. As $Q_1$ is small, $V\rightarrow0$, the particle stays at the bottom of the potential and can not pass the barrier. We also find, for $Q_2=5$, $V<0$ as $Q_1>800$, so the current reversal phenomenon appear with increasing $Q_1$.

\begin{figure}
\center{
\includegraphics[height=6cm,width=8cm]{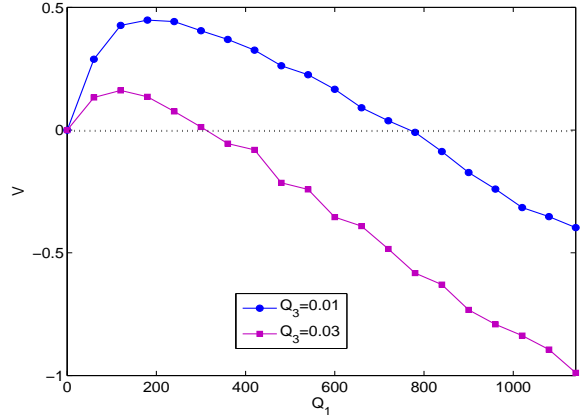}
\caption{Average velocity $V$ as a function of $Q_1$ with different $Q_3$. The other parameters are $Q_2=5$, $\lambda=0.9$, $v_0=2$ and $\tau_1=\tau_2=\tau_3=1$.}
\label{Fig3}}
\end{figure}
Figure \ref{Fig3} shows the dependence of $V$ on $Q_1$ with different self-propelled angle noise intensity $Q_3$. Just like Fig.(\ref{Fig2}), $V$ has a maximum with increasing $Q_1$.  We find the peak of the curve moves to left with decreasing $Q_3$, and the smaller $Q_3$, the higher of the peak is.

\begin{figure}
\center{
\includegraphics[height=6cm,width=8cm]{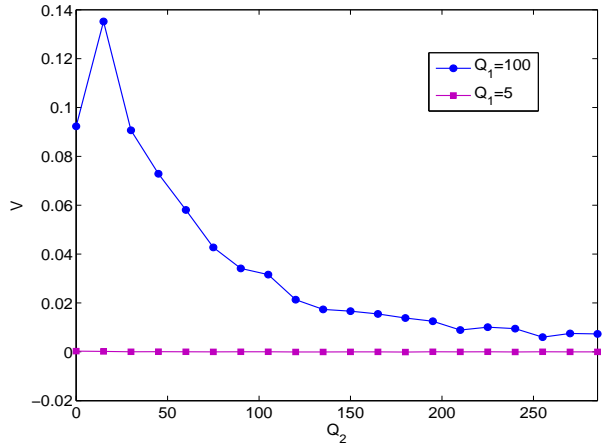}
\caption{Average velocity $V$ as a function of the $y$ direction noise intensity $Q_2$ with different $Q_1$. The other parameters are $Q_3=0.03$, $\lambda=0.9$, $v_0=2$ and $\tau_1=\tau_2=\tau_3=1$.}
\label{Fig4}}
\end{figure}
Average velocity $V$ as a function of $y$ direction noise intensity $Q_2$ with different $Q_1$ is shown in Fig.(\ref{Fig4}). $V$ has a marked peak with increasing $Q_2$ as $Q_1$ is large($Q_1=100$), and $V$ has always been zero with increasing $Q_2$ as $Q_1$ small($Q_1=5$). So $x$ direction noise $Q_1$ can induce $x$ direction particles transport, and an optimal value of $y$ direction noise $Q_2$ appear at which $V$ takes its maximal value. As $Q_1$ is small, the particle stays at the bottom of the potential and can not transport the potential regardless of $Q_2$ large or small. $x$ direction noise can induce $x$ direction particles transport, and $y$ direction noise have some effects on the particles transport phenomenon, but $y$ direction noise can not induce $x$ direction particles transport as $x$ direction noise is small.

\begin{figure}
\center{
\includegraphics[height=6cm,width=8cm]{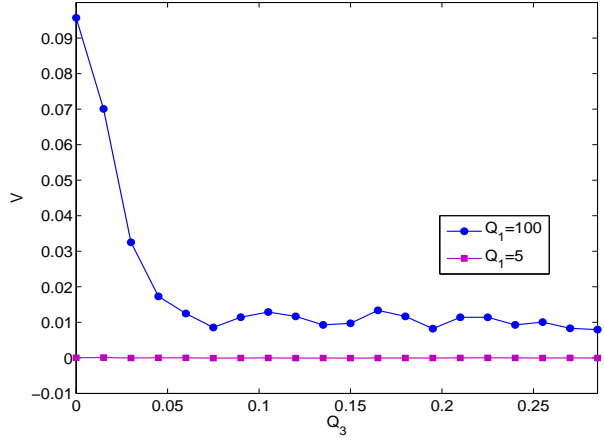}
\caption{Average velocity $V$ as a function of the noise intensity $Q_3$ with different $Q_1$. The other parameters are $Q_2=100$, $\lambda=0.9$, $v_0=2$ and $\tau_1=\tau_2=\tau_3=1$.}
\label{Fig5}}
\end{figure}
Figure \ref{Fig5} shows the dependence of $V$ on the self-propelled angle noise intensity $Q_3$ with different $Q_1$. As $Q_1=100$, we find $V$  appears periodic fluctuations and the amplitude becomes smaller and smaller with increasing $Q_3$. $V$ has always been zero with increasing $Q_3$ as $Q_1$ is small($Q_1=5$). So as $Q_1$ is large, small $Q_3$ should maximize the $x$ direction velocity $V$, and  self-propelled angle noise can not induce $x$ direction particles transport itself as $Q_1$ is small. From Fig.(\ref{Fig4}) and Fig.(\ref{Fig5}), we find $Q_2$ and $Q_3$ can not induce particles transport themselves, but they have great effects on $Q_1$ induced $x$ direction particles transport.

\begin{figure}
\center{
\includegraphics[height=6cm,width=8cm]{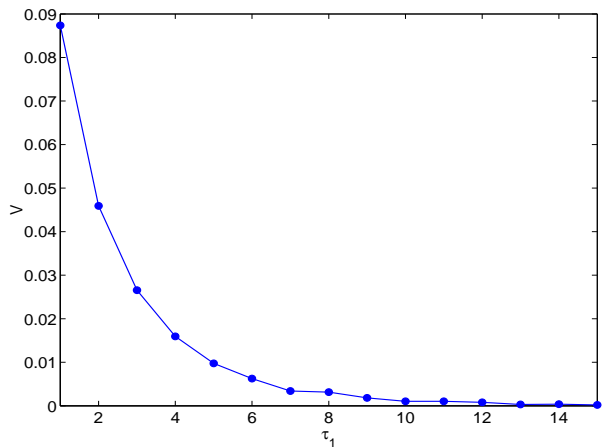}
\caption{Average velocity $V$ as a function of the self-correlation time $\tau_1$. The other parameters are $Q_1=Q_2=100$, $Q_3=0.03$, $\lambda=0.9$, $v_0=2$  and $\tau_2=\tau_3=1$.}\label{Fig6}
}
\end{figure}
Figure \ref{Fig6} shows the dependence of $V$ on the the self-correlation time $\tau_1$. It is found $V$ decreases monotonically with increasing $\tau_1$. So large $\tau_1$ cause the inhibition of the particles transport.

\begin{figure}
\center{
\includegraphics[height=6cm,width=8cm]{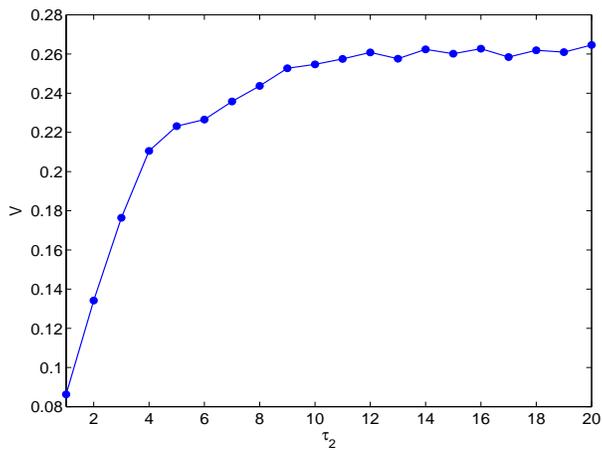}
\caption{Average velocity $V$ as a function of the self-correlation time $\tau_2$. The other parameters are $Q_1=Q_2=100$, $Q_3=0.03$,$\lambda=0.9$, $v_0=2$  and $\tau_1=\tau_3=1$.}\label{Fig7}
}
\end{figure}
Average velocity $V$ as a function of self-correlation time $\tau_2$ is shown in Fig.(\ref{Fig7}). Unlike Fig.(\ref{Fig5}), we find $V$ increase with increasing $\tau_2$, so large $\tau_2$  is better for particles transport.

\begin{figure}
\center{
\includegraphics[height=6cm,width=8cm]{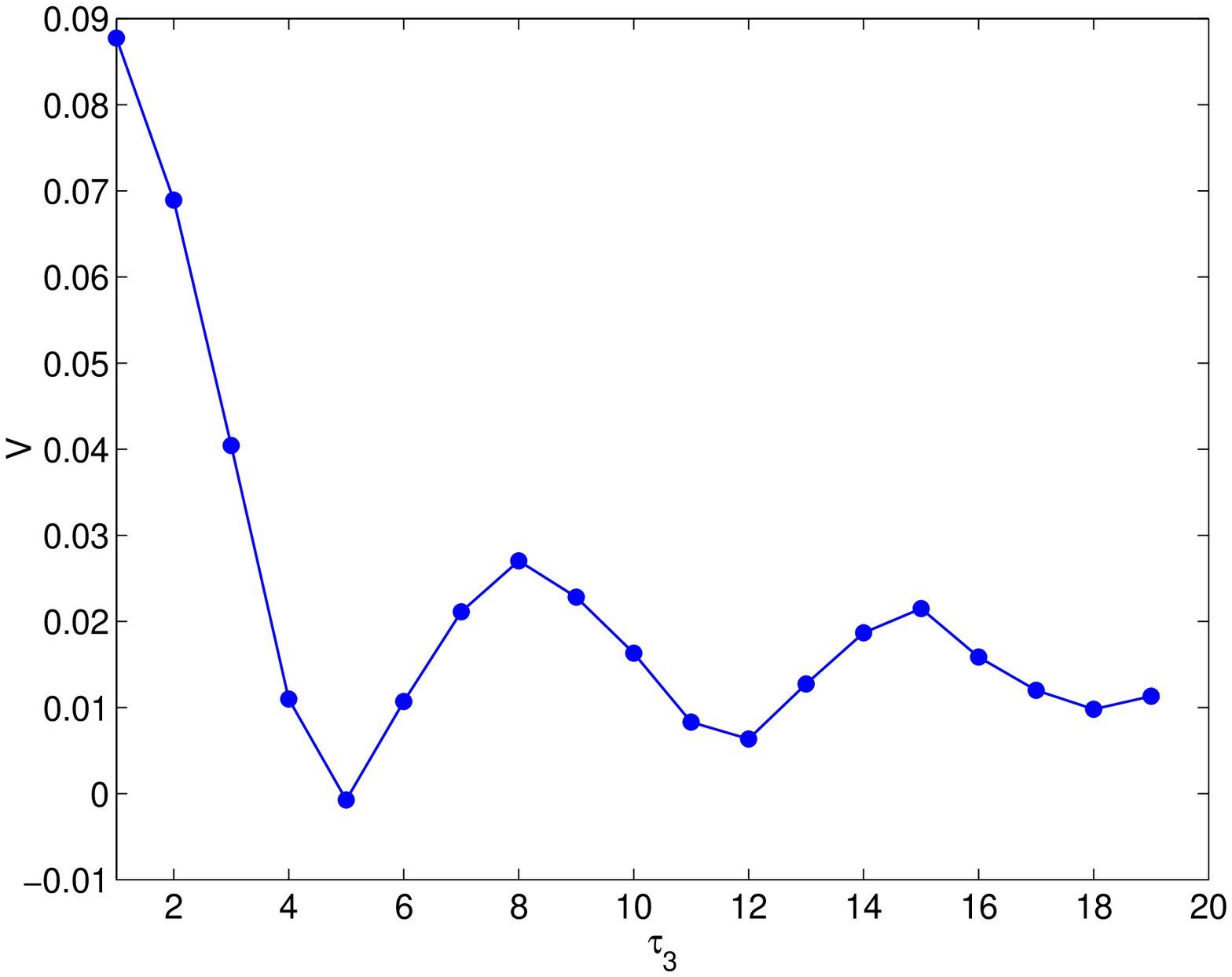}
\caption{Average velocity $V$ as a function of the self-correlation time $\tau_3$. The other parameters are $Q_1=Q_2=100$, $Q_3=0.03$, and $\tau_1=\tau_2=1$.}\label{Fig8}
}
\end{figure}
Figure \ref{Fig8} depicts average velocity $V$ as a function of the self-correlation time $\tau_3$. We can find $V$ appear periodic fluctuations with increasing $\tau_3$, and the amplitude becomes smaller and smaller with increasing $\tau_3$.

\begin{figure}
\center{
\includegraphics[height=6cm,width=8cm]{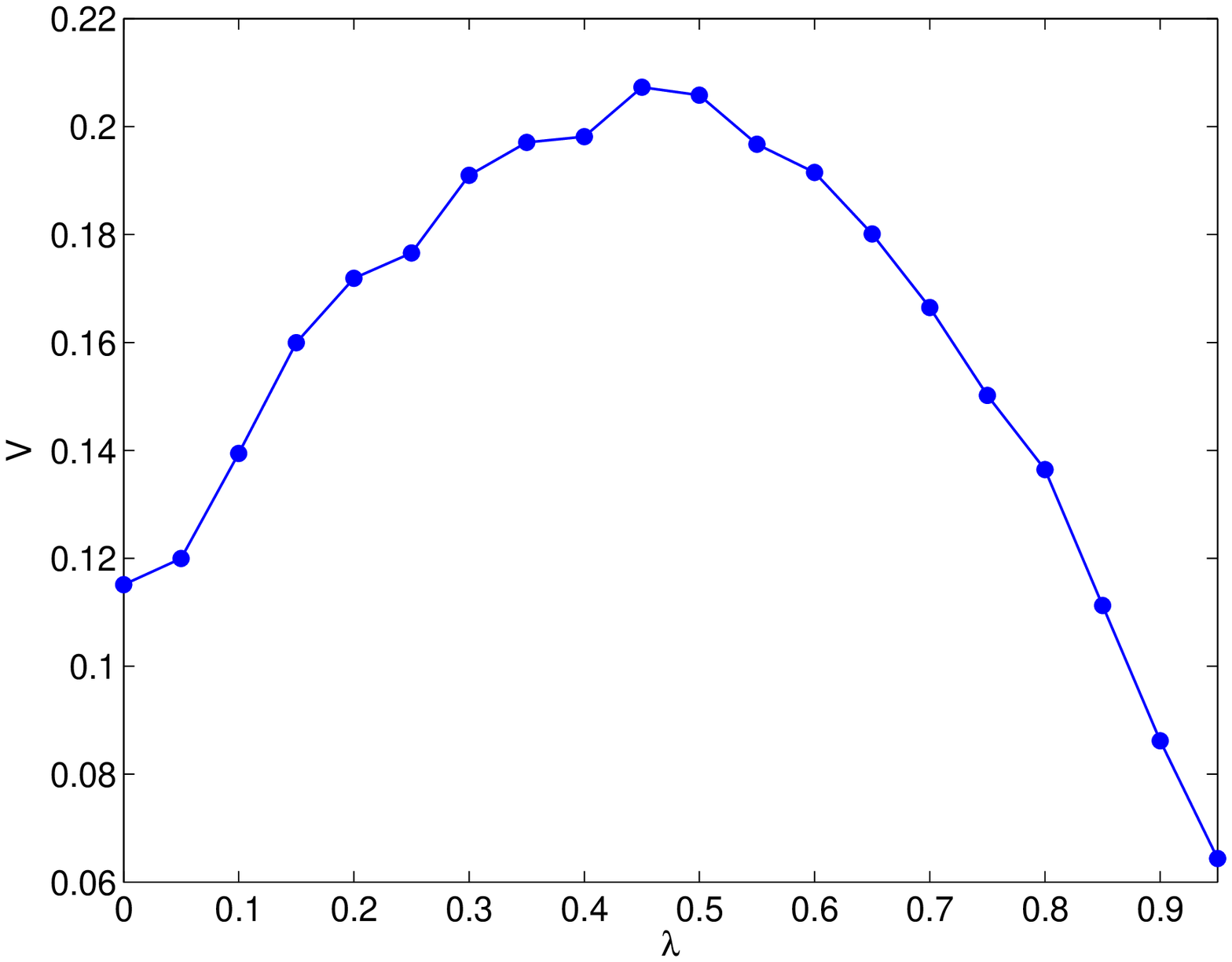}
\caption{Average velocity $V$ as a function of the modulation constant $\lambda$. The other parameters are $Q_1=Q_2=100$, $Q_3=0.03$, $v_0=2$ and $\tau_1=\tau_2=1$.}\label{Fig9}
}
\end{figure}
Average velocity $V$ as a function of the modulation constant $\lambda$ is shown in Fig.\ref{Fig9}. As the modulation constant $\lambda$ increases, the average velocity first increases,
and then decreases near $\lambda=0.5$. There exists an optimal value of $\lambda$ at which the average velocity $V$ takes its maximal value. When there is no coupling between the $x$ direction potential and the $y$ direction potential($\lambda=0$), the average velocity $V\neq0$. The results differ from the results of Ref.\cite{b.i}, the major reason is  the independent of noise $\xi_1$ and $\xi_2$ in our study, and $\xi_1$ and $\xi_2$ have different effects on the system.

\begin{figure}
\center{
\includegraphics[height=6cm,width=8cm]{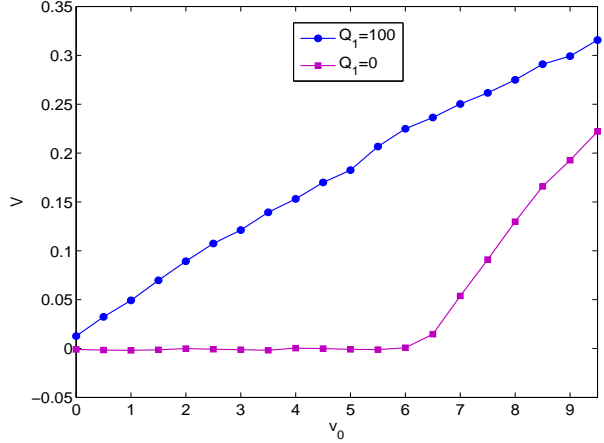}
\caption{Average velocity $V$ as a function of the self-propelled velocity $v_0$ with different $Q_1$. The other parameters are $Q_2=100$, $Q_3=0.03$, and $\tau_1=\tau_2=\tau_3=1$ . }\label{Fig10}
}
\end{figure}
Figure \ref{Fig10} depicts $V$ as a function of the self-propelled velocity $v_0$ with different $Q_1$. As $Q_1=100$, we find $V\neq0$ as $v_0=0$, so proper $\xi_1$ can induce particles transport itself  whether the self-propelled velocity exists or not. As $Q_1=0$, it is found that the average velocity is zero until $v_0>6$, the reason is the value of the $x$ direction force for the potential is $|\hat{F}_x|=|-\frac{\partial{\hat{U}(x,y)}}{\partial{\hat{x}}}|=|2\pi\cos(2\pi \hat{x})|\leq2\pi$. Therefore, in order to pass through the channel, the particle needs a nonzero $v_0$ to move against the max force $F_{x(max)}=2\pi$. This result is the same as the result of Ref.\cite{b.i}.

\section{\label{label4}Conclusions}
In this paper, we numerically studied the transport of self-propelled particles in a two-dimensional potential with colored noise. We find that the average velocity $V$ has a maximum with increasing $x$ direction diffusion constant $Q_1$. Large $Q_1$ can induce current reversal phenomenon. $y$ direction diffusion constant and the rotational diffusion rate $Q_3$ can not induce $x$ direction particles transports themselves, but they have great effects on $x$ direction average velocity $V$. $V$ has a maximum with increasing $Q_2$ and appear periodic fluctuations with increasing $Q_3$. $V$ decreases with increasing $\tau_1$(self-correlation time of $x$ direction noise) and increases with increasing $\tau_2$(self-correlation time of $y$ direction noise). $V$ appear periodic fluctuations with increasing self-correlation time $\tau_3$.

\section{Acknowledgments}
Project supported by Natural Science Foundation of Anhui Province(Grant No:1408085QA11) and the National Natural Science Foundation of China (No:11404005).

\bibliographystyle{elsarticle-num}

\end{document}